\documentclass[12pt]{article}
\usepackage{amssymb}
\usepackage{graphicx}
\usepackage[cp1251]{inputenc}
\usepackage{rotating}

\begin{document}
 \noindent {\footnotesize\it Astronomy Letters, 2024, Vol. 50, No. 4, pp. 239--249}
 \newcommand{\dif}{\textrm{d}}

 \noindent
 \begin{tabular}{llllllllllllllllllllllllllllllllllllllllllllll}
 & & & & & & & & & & & & & & & & & & & & & & & & & & & & & & & & & & & & & &\\\hline\hline
 \end{tabular}

  \vskip 0.5cm
  \bigskip
 \bigskip
\centerline{\large\bf  Estimation of the Kinematic Age of the $\beta$ Pictoris Moving Group}
\centerline{\large\bf from Up-To-Date Data}

 \bigskip
 \bigskip
  \centerline { %  DOI: 10.1134/S1063773724700117
   V. V. Bobylev\footnote [1]{bob-v-vzz@rambler.ru},  A. T. Bajkova}
 \bigskip
 \centerline{\small\it Pulkovo Astronomical Observatory, Russian Academy of Sciences, St. Petersburg, 196140 Russia}
 \bigskip
 \bigskip
{Abstract---The kinematics of about 40 single stars belonging to the $\beta$~Pictoris moving group is studied.
The age of the $\beta$~Pictoris moving group is estimated from these stars with ground-based line-of-sight
velocity determinations by two methods. Both estimates are kinematic. In the first method we considered
the traceback trajectories of the stars, giving an estimate of $t = 13.2\pm1.4$ Myr. In the second method, by
analyzing the instantaneous velocities of the stars, we show that there is an expansion of the stellar system
occurring in the Galactic $xy$ plane. Based on this effect, we find the time interval elapsed from the beginning
of the expansion of the $\beta$~Pictoris moving group to the present day, $t = 20\pm2$ Myr.
 }

\bigskip
\section*{INTRODUCTION}
The $\beta$~Pictoris moving group consists of young (with an age $\sim$20~Myr) stars that are distributed around the Sun within $\sim$50 pc. Owing to the close proximity of these stars to the Sun, their comprehensive study is of great importance for solving various stellar-astronomy problems. Most of the group members are low-mass K- and M-type stars; on the Hertzsprung--Russell diagram they occupy a region typical for pre-main-sequence stars. There is still lithium in their atmospheres whose abundance analysis underlies one of the methods for estimating the ages of star clusters and associations. The star $\beta$~Pictoris proper is a fairly bright A-type star. It has been established by direct observations that the star is surrounded by a debris disk, several belts of planetesimals, and two exoplanets orbit it, $\beta$~Pict b and $\beta$~Pict c. The dynamical interaction of these exoplanets with the disk, planetesimals, and exocomets was simulated in the interesting paper of Beust et al. (2024). 

Open star clusters (OSCs) contain $\sim$103 members and are gravitationally bound systems on long time scales, $\sim$2--4 Gyr. Many of the OSCs, for example, the Pleiades or the Hyades, can be observed in the sky as a distinct compact clump of stars. In contrast to OSCs, moving stellar groups contain a small number of members (several tens), they do not forma distinct clump in the sky, and they are identified by the common space motion of stars, metallicity, age, and other common characteristics. Several such structures of various ages are known to date. For example, these include the Ursa Major, Castor, or $\zeta$~Herculis moving stellar groups.

Much work in searching for and analyzing moving stellar groups was done at the time by O. Eggen. For example, he described the Sirius kinematic group (Eggen 1960). Agekyan and Orlov (1984) analyzed several moving stellar groups based on data from the Catalog of Nearby Stars (Glieze 1969). The star $\beta$~Pictoris (Glieze 219) turned out to be a member of even two groups.

The Hipparcos (1997) catalogue exerted a revolutionary influence on the search for and analysis of moving stellar groups. For example, beginning with the papers of Barrado y Navascu\'es et al. (1999) and Zuckerman et al. (2001) based on Hipparcos data, the $\beta$~Pictoris moving groupwas described in a form close to the currently adopted one. The list of stellar group candidates was supplemented by Torres et al. (2006),
Schlieder et al. (2010, 2012), Kiss et al. (2011), Malo et al. (2014), Riedel et al. (2014), and Shkolnik et al. (2017) using the results of various observational programs. As a result, for a number of group members the most important kinematic characteristics, in particular, their line-of-sight velocities, were refined and redetermined.

A new surge of interest in studying moving stellar groups was caused by the publication of highly accurate data from several versions of the Gaia
catalog (Prusti et al. 2016). In particular, up-todate results of the analysis of the $\beta$~Pictoris moving group are presented in Crundall et al. (2019), Miret-Roig et al. (2020), Couture et al. (2023), and Lee et al. (2024).

Various methods, such as isochrone fitting, lithium abundance analysis, kinematic methods, etc., have been used to estimate the age of the 
$\beta$~Pictoris moving group. Many of these estimates are presented in Lee et al. (2024), where poor agreement of the results obtained by various methods can be noticed.

The goal of this paper is to estimate the kinematic age of the $\beta$~Pictoris moving group. For this purpose, we use data from the Gaia DR3 catalog (Vallenari et al. 2023). The sample is based on exclusively single stars, according to the classification of Lee et al. (2024). The method consists in tracing back the orbits of stars on a given time interval and estimating the time when the stellar group had a minimum spatial size.

\section*{METHOD}
We use a rectangular coordinate system centered on the Sun, where the $x$ axis is directed towards the galactic center, the $y$ axis~--- towards galactic rotation and the $z$ axis~--- to the north pole of the Galaxy. Then $x=r\cos l\cos b,$ $y=r\sin l\cos b$ and $z=r\sin b,$ where $r=1/\pi$ is the heliocentric distance of the star in the kpc, which we calculate through the parallax of the star $\pi$ in mas (milliseconds of arc).

The radial velocity $V_r$ and two tangential velocity projections $V_l=4.74r\mu_l\cos b$ and $V_b=4.74r\mu_b$  along the galactic longitude $l$ and latitude $b$, respectively, expressed in km s$^{-1}$ are known from observations. Here, the coefficient 4.74 is the ratio of the number of kilometers in an astronomical unit to the number of seconds in a tropical year. The components of proper motion $\mu_l\cos b$ and $\mu_b$ are expressed in mas yr$^{-1}$.

Through the components $V_r, V_l, V_b,$ the velocities $U,V,W,$ are calculated, where the velocity $U$ is directed from the Sun to the center of the Galaxy, $V$ in the direction of rotation of the Galaxy and $W$ to the north galactic pole:
 \begin{equation}
 \begin{array}{lll}
 U=V_r\cos l\cos b-V_l\sin l-V_b\cos l\sin b,\\
 V=V_r\sin l\cos b+V_l\cos l-V_b\sin l\sin b,\\
 W=V_r\sin b                +V_b\cos b.
 \label{UVW}
 \end{array}
 \end{equation}
Thus, the velocities $U,V,W$ are directed along the corresponding coordinate axes $x,y,z$.

TTo construct the stellar orbits in a coordinate system  rotating around the Galactic center, we use the epicyclic approximation~(Lindblad 1927):
 \begin{equation}
 \renewcommand{\arraystretch}{1.8}
 \begin{array}{lll}\displaystyle
 x(t)= x_0+{U_0\over \displaystyle \kappa}\sin(\kappa t)+{\displaystyle V_0\over \displaystyle 2B}(1-\cos(\kappa t)),  \\
 y(t)= y_0+2A \biggl(x_0+{\displaystyle V_0\over\displaystyle 2B}\biggr) t
       -{\displaystyle \Omega_0\over \displaystyle B\kappa} V_0\sin(\kappa t)
       +{\displaystyle 2\Omega_0\over \displaystyle \kappa^2} U_0(1-\cos(\kappa t)),\\
 z(t)= {\displaystyle W_0\over \displaystyle \nu} \sin(\nu t)+z_0\cos(\nu t),
 \label{EQ-Epiciclic}
 \end{array}
 \end{equation}
where $t$ is time in million years (based on the ratio pc/Myr=0.978 km s$^{-1}$),
$A$ and $B$ are Oort constants; $\kappa=\sqrt{-4\Omega_0 B}$ is epicyclic frequency; $\Omega_0$ is angular velocity of galactic rotation of the local standard of rest, $\Omega_0=A-B$; $\nu=\sqrt{4\pi G\rho_0}$ is frequency of vertical oscillations, where $G$ is gravitational constant, and $\rho_0$ is stellar density in the near-solar neighborhood.

The parameters $x_0,y_0,z_0$ and $U_0,V_0,W_0$ in the system of equations~(2) denote the current positions and velocities of the stars, respectively. The elevation of the Sun above the galactic plane $h_\odot$ is assumed to be $16$~pc according to the work~(Bobylev and Bajkova 2016). The velocities $U,V,W$ are calculated relative to the local standard of rest using the values $(U_\odot,V_\odot,W_\odot)=(11.1,12.2,7.3)$~km s$^{-1}$ from Sch\"onrich et al. (2010. We took $\rho_0=0.1~M_\odot/$pc$^3$~(Holmberg and
Flinn 2004), which gives $\nu=74$~km s$^{-1}$ kpc$^{-1}$. We use the following values of the Oort constants $A=16.9$~km s$^{-1}$ kpc$^{-1}$ and $B=-13.5$~km s$^{-1}$ kpc$^{-1}$, close to modern estimates. An overview of such estimates can be found, for example, in Krisanova
et al. (2020).

%%%%%%%%%%%%%%%%%%%%%%%%%%%%%%%%%%%%%%%%%%%%%%%%%%%%%%%%%%%%% t-1
  \begin{table}[p]
  \caption[]{\small
Single stars of the $\beta$ Pictoris moving group, $\mu^*_\alpha=\mu_\alpha\cos\delta$
 }
  \begin{center}  \label{Table-1}    \small
  \begin{tabular}{|r|r|r|r|r|r|r|}\hline
 Gaia DR3 & $\alpha,$ deg & $\delta,$~deg & $\pi\pm\sigma,$~mas & $\mu^*_\alpha\pm\sigma,$~mas/yr & 
 $\mu_\delta\pm\sigma,$~mas/yr \\\hline

66245408072670336   &  59.39 & $ 24.75$ & $ 14.55\pm .02 $ & $ 34.34\pm .03 $ & $ -46.52\pm .02 $ \\
2901786974419551488 &  82.44 & $-32.65$ & $ 33.60\pm .02 $ & $ 15.27\pm .02 $ & $  10.77\pm .02 $ \\
3238965099979863296 &  76.55 & $  4.66$ & $ 36.19\pm .03 $ & $ 30.16\pm .03 $ & $ -89.86\pm .02 $ \\
3231945508509506176 &  74.90 & $  1.78$ & $ 40.99\pm .01 $ & $ 39.13\pm .01 $ & $ -94.90\pm .01 $ \\
4764027962957023104 &  75.20 & $-57.26$ & $ 37.21\pm .01 $ & $ 35.39\pm .01 $ & $  74.11\pm .02 $ \\
5412403269717562240 & 146.62 & $-44.96$ & $ 21.44\pm .03 $ & $-78.26\pm .03 $ & $   9.26\pm .03 $ \\
5963633872326630272 & 255.67 & $-45.37$ & $ 31.30\pm .02 $ & $-20.10\pm .02 $ & $-137.85\pm .02 $ \\
5924485966955008896 & 262.48 & $-54.26$ & $ 14.79\pm .01 $ & $ -5.49\pm .01 $ & $ -63.44\pm .01 $ \\
4067828843907821824 & 268.01 & $-23.97$ & $ 15.67\pm .02 $ & $   .16\pm .03 $ & $ -52.50\pm .02 $ \\
4050178830427649024 & 271.07 & $-30.31$ & $ 18.15\pm .02 $ & $  3.42\pm .02 $ & $ -65.22\pm .02 $ \\
6648834361774839040 & 271.48 & $-57.08$ & $ 17.71\pm .02 $ & $   .89\pm .02 $ & $ -72.95\pm .02 $ \\
6649786646225001984 & 280.52 & $-55.90$ & $ 19.36\pm .02 $ & $ 11.12\pm .02 $ & $ -78.05\pm .01 $ \\
6649788119394186112 & 280.53 & $-55.91$ & $ 19.44\pm .02 $ & $ 12.01\pm .02 $ & $ -79.07\pm .01 $ \\
6631685008336771072 & 281.72 & $-62.18$ & $ 19.72\pm .02 $ & $ 13.24\pm .02 $ & $ -80.28\pm .02 $ \\
6736232346363422336 & 282.69 & $-31.80$ & $ 20.22\pm .01 $ & $ 17.27\pm .02 $ & $ -72.34\pm .01 $ \\
6655168686921108864 & 283.27 & $-50.18$ & $ 21.16\pm .02 $ & $ 16.27\pm .02 $ & $ -85.52\pm .02 $ \\
6663346029775435264 & 290.91 & $-46.11$ & $ 14.03\pm .02 $ & $ 18.07\pm .02 $ & $ -57.25\pm .01 $ \\
6764421281858414208 & 292.52 & $-29.66$ & $ 16.60\pm .02 $ & $ 23.67\pm .03 $ & $ -59.61\pm .02 $ \\
6754492966739292928 & 297.07 & $-27.34$ & $ 15.47\pm .02 $ & $ 25.15\pm .02 $ & $ -53.38\pm .01 $ \\
6747467224874108288 & 299.02 & $-32.13$ & $ 19.49\pm .02 $ & $ 33.60\pm .02 $ & $ -68.53\pm .01 $ \\
6747106443324127488 & 300.41 & $-33.22$ & $ 16.68\pm .02 $ & $ 29.23\pm .02 $ & $ -61.39\pm .01 $ \\
6700649538727351040 & 301.49 & $-32.28$ & $ 20.18\pm .03 $ & $ 38.44\pm .03 $ & $ -70.45\pm .02 $ \\
6794047652729201024 & 311.29 & $-31.34$ & $102.94\pm .02 $ & $281.32\pm .02 $ & $-360.15\pm .02 $ \\
6833292181958100224 & 317.52 & $-19.33$ & $ 30.90\pm .03 $ & $ 90.61\pm .03 $ & $ -91.00\pm .02 $ \\
6835588645136005504 & 320.03 & $-16.76$ & $ 20.72\pm .02 $ & $ 59.81\pm .02 $ & $ -58.13\pm .02 $ \\
6382640367603744128 & 340.71 & $-71.71$ & $ 27.23\pm .01 $ & $ 94.85\pm .01 $ & $ -52.38\pm .01 $ \\
2433191886212246784 & 353.13 & $-12.26$ & $ 36.43\pm .02 $ & $139.63\pm .02 $ & $ -82.07\pm .02 $ \\
87555176071871744   &  36.07 & $ 20.53$ & $ 14.13\pm .07 $ & $ 47.04\pm .08 $ & $ -39.79\pm .07 $ \\
68012529415816832   &  53.76 & $ 23.71$ & $ 19.72\pm .09 $ & $ 51.48\pm .11 $ & $ -62.85\pm .08 $ \\
5266270443442455040 &  94.62 & $-72.04$ & $ 25.57\pm .01 $ & $ -7.71\pm .02 $ & $  74.41\pm .01 $ \\
6414282147589248000 & 272.28 & $-76.22$ & $ 36.66\pm .02 $ & $  7.20\pm .02 $ & $-150.57\pm .02 $ \\
4071532308311834496 & 281.81 & $-28.15$ & $ 16.69\pm .04 $ & $ 14.38\pm .05 $ & $ -60.94\pm .04 $ \\
6850555648387276544 & 302.16 & $-25.76$ & $ 17.85\pm .04 $ & $ 35.14\pm .03 $ & $ -59.42\pm .02 $ \\
6801191424589717888 & 317.63 & $-27.18$ & $ 24.84\pm .06 $ & $ 70.16\pm .05 $ & $ -76.06\pm .04 $ \\
6801191355870240768 & 317.63 & $-27.18$ & $ 24.76\pm .03 $ & $ 68.01\pm .03 $ & $ -75.68\pm .02 $ \\
6608255235884536320 & 338.45 & $-29.84$ & $ 19.19\pm .10 $ & $ 65.15\pm .09 $ & $ -45.92\pm .09 $ \\
2324205785406060928 & 353.96 & $-34.03$ & $ 26.76\pm .04 $ & $101.72\pm .04 $ & $ -50.22\pm .04 $ \\
2315849737553379840 &   7.06 & $-32.47$ & $ 28.56\pm .02 $ & $112.01\pm .02 $ & $ -44.58\pm .03 $ \\
2357025657739386624 &  12.11 & $-18.79$ & $ 19.42\pm .03 $ & $ 73.28\pm .03 $ & $ -47.15\pm .03 $ \\
5177677603263978880 &  41.28 & $ -7.14$ & $ 14.66\pm .02 $ & $ 45.04\pm .03 $ & $ -37.59\pm .03 $ \\
6603693808817829760 & 341.25 & $-33.26$ & $ 48.00\pm .03 $ & $176.82\pm .03 $ & $-120.88\pm .02 $ \\
 \hline
 \end{tabular}\end{center}
 \end{table}
%%%%%%%%%%%% t
%%%%%%%%%%%%%%%%%%%%%%%%%%%%%%%%%%%%%%%%%%%%%%%%%%%%%%%%%%%%% t-2
  \begin{table}[p]
  \caption[]{\small
Line-of-sight velocities of the selected single stars
 }
  \begin{center}  \label{Table-2}    \small
  \begin{tabular}{|r|r|r|r|r|c|c|}\hline
 Gaia DR3 & $\alpha$ ~~ & $\delta$ ~~ &
  $(V_r\pm\sigma)_{Gaia}$ & $(V_r\pm\sigma)_{SIMBAD}$ \\
 & deg & deg & km s$^{-1}$ ~~~~ & km s$^{-1}$ ~~~~ \\\hline

66245408072670336   &  59.39 & $ 24.75$ & $ 11.85\pm~0.66 $ & $ 14.30\pm0.01 $ \\
2901786974419551488 &  82.44 & $-32.65$ & $ 23.39\pm18.39 $ & $ 22.00\pm0.60 $ \\
3238965099979863296 &  76.55 & $  4.66$ & $ 20.13\pm~3.73 $ & $ 18.80\pm2.40 $ \\
3231945508509506176 &  74.90 & $  1.78$ & $ 18.54\pm~0.25 $ & $ 19.16\pm0.01 $ \\
4764027962957023104 &  75.20 & $-57.26$ & $ 18.54\pm~0.22 $ & $ 19.16\pm0.01 $ \\
5412403269717562240 & 146.62 & $-44.96$ & $ 15.84\pm~3.39 $ & $ 15.69\pm1.52 $ \\
5963633872326630272 & 255.67 & $-45.37$ & $ -3.65\pm~0.43 $ & $ -2.63\pm0.18 $ \\
5924485966955008896 & 262.48 & $-54.26$ & $ -1.58\pm~1.50 $ & $  3.44\pm0.09 $ \\
4067828843907821824 & 268.01 & $-23.97$ & $-10.74\pm~3.18 $ & $-10.23\pm1.82 $ \\
4050178830427649024 & 271.07 & $-30.31$ & $ -7.19\pm~1.54 $ & $ -7.41\pm0.24 $ \\
6648834361774839040 & 271.48 & $-57.08$ & $ -1.17\pm~1.59 $ & $  0.65\pm0.21 $ \\
6649786646225001984 & 280.52 & $-55.90$ & $  3.12\pm26.81 $ & $  0.43\pm0.85 $ \\
6649788119394186112 & 280.53 & $-55.91$ & $  0.20\pm~1.28 $ & $  1.17\pm0.17 $ \\
6631685008336771072 & 281.72 & $-62.18$ & $  1.45\pm~0.46 $ & $  1.72\pm0.01 $ \\
6736232346363422336 & 282.69 & $-31.80$ & $-12.05\pm~2.24 $ & $ -8.81\pm0.20 $ \\
6655168686921108864 & 283.27 & $-50.18$ & $ -3.59\pm~1.55 $ & $ -4.20\pm0.20 $ \\
6663346029775435264 & 290.91 & $-46.11$ & $ -1.29\pm~0.49 $ & $ -0.31\pm0.49 $ \\
6764421281858414208 & 292.52 & $-29.66$ & $ -3.33\pm~7.13 $ & $ -5.17\pm0.95 $ \\
6754492966739292928 & 297.07 & $-27.34$ & $ -6.20\pm~1.26 $ & $ -6.26\pm0.16 $ \\
6747467224874108288 & 299.02 & $-32.13$ & $ -6.57\pm~0.36 $ & $ -6.15\pm0.05 $ \\
6747106443324127488 & 300.41 & $-33.22$ & $ -4.13\pm~0.46 $ & $ -4.36\pm0.08 $ \\
6700649538727351040 & 301.49 & $-32.28$ & $ -6.55\pm~0.61 $ & $ -5.10\pm1.30 $ \\
6794047652729201024 & 311.29 & $-31.34$ & $ -6.90\pm~0.37 $ & $ -4.71\pm0.01 $ \\
6833292181958100224 & 317.52 & $-19.33$ & $ -6.43\pm~0.88 $ & $ -6.14\pm0.01 $ \\
6835588645136005504 & 320.03 & $-16.76$ & $ -6.07\pm~2.08 $ & $ -5.10\pm0.62 $ \\
6382640367603744128 & 340.71 & $-71.71$ & $  7.02\pm~0.21 $ & $  7.99\pm0.02 $ \\
2433191886212246784 & 353.13 & $-12.26$ & $ -0.71\pm~0.69 $ & $  0.83\pm0.29 $ \\
  87555176071871744 &  36.07 & $ 20.53$ & $   $ & $  8.62\pm1.22 $ \\
  68012529415816832 &  53.76 & $ 23.71$ & $   $ & $ 15.50\pm1.70 $ \\
5266270443442455040 &  94.62 & $-72.04$ & $   $ & $ 16.10\pm0.01 $ \\
6414282147589248000 & 272.28 & $-76.22$ & $   $ & $  6.98\pm0.38 $ \\
4071532308311834496 & 281.81 & $-28.15$ & $   $ & $ -7.46\pm1.65 $ \\
6850555648387276544 & 302.16 & $-25.76$ & $   $ & $ -5.74\pm1.57 $ \\
6801191424589717888 & 317.63 & $-27.18$ & $   $ & $ -3.80\pm0.40 $ \\
6801191355870240768 & 317.63 & $-27.18$ & $   $ & $ -4.21\pm0.33 $ \\
6608255235884536320 & 338.45 & $-29.84$ & $   $ & $ -1.94\pm0.30 $ \\
2324205785406060928 & 353.96 & $-34.03$ & $   $ & $  5.90\pm0.78 $ \\

2315849737553379840 &   7.06 & $-32.47$ & $ 12.08\pm~2.69$ & $  6.79\pm2.66 $ \\
2357025657739386624 &  12.11 & $-18.79$ & $ 11.81\pm~3.42$ & $  7.21\pm0.68 $ \\
5177677603263978880 &  41.28 & $ -7.14$ & $  4.39\pm~3.45$ & $ 11.36\pm2.29 $ \\
6603693808817829760 & 341.25 & $-33.26$ & $ -4.65\pm~3.83$ & $  1.84\pm0.02 $ \\
\hline
 \end{tabular}\end{center}
 \end{table}
%%%%%%%%%%%% t2

\section*{DATA}
The stars that were classified as single ones in Lee et al. (2024), where a detailed analysis of the $\beta$~Pictoris moving group candidates was performed, form the basis for our sample. The original sample of candidates in the paper of these authors included 415 stars from the Gaia DR3 catalog, for 99 of which the line-of-sight velocities from this catalog were measured. According to Lee et al. (2024), the final list of probable $\beta$~Pictoris moving group candidates contains 86 stars (single, confirmed binary, and unresolved binary ones).

From this list we selected single stars and produced two samples. The first sample included 31 stars with the trigonometric parallaxes, proper motions, and line-of-sight velocities from the Gaia DR3 catalog. The second sample included 41 stars with the
parallaxes and proper motions from the Gaia DR3 catalog and the line-of-sight velocities from the SIMBAD
electronic database,1 where the corresponding references can be found.

The choice of precisely single stars is fairly obvious. After all, the orbital motion of the components of a binary or multiple system around a common center of mass can distort significantly the results of our kinematic analysis. Of course, this reasoning refers to the case where the instantaneous stellar velocities were measured, while the orbital parameters of a binary or multiple system were not determined.

The coordinates, trigonometric parallaxes, and proper motions of the selected stars taken from the
Gaia DR3 catalog are given in Table 1. Two versions of the line-of-sight velocities are given in Table 2.

%%%%%%%%%%%%%%%%%%%%%%%%%%%%%%%%%%%%%%%%%%%%%%%%%%%%%%%%%%%%% F1:
\begin{figure}[t]
{ \begin{center}
  \includegraphics[width=0.9\textwidth]{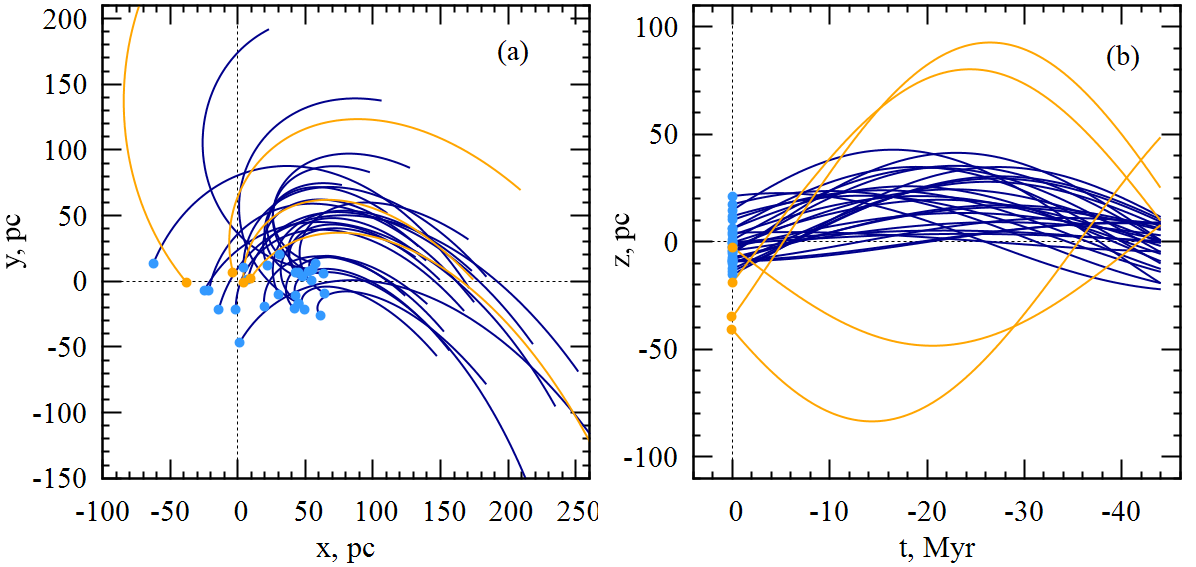}
  \caption{
 (a) The distribution of 31 members of the $\beta$~Pictoris moving group in projection onto the Galactic $xy$ plane and their
traceback trajectories, (b) the vertical distribution of these stars and their traceback trajectories, the trajectories are traced back
on an interval of 45~Myr, the orange color marks the trajectories of the four stars discarded when calculating the kinematic
center of this stellar group. We took all of the data for these stars from the Gaia DR3 catalog.}
 \label{f1-DR3}
\end{center}}
\end{figure}
%%%%%%%%%%%%%%%%%%%%  f1
%%%%%%%%%%%%%%%%%%%%%%%%%%%%%%%%%%%%%%%%%%%%%%%%%%%%%%%%%%%%% F2:
\begin{figure}[t]
{ \begin{center}
   \includegraphics[width=0.9\textwidth]{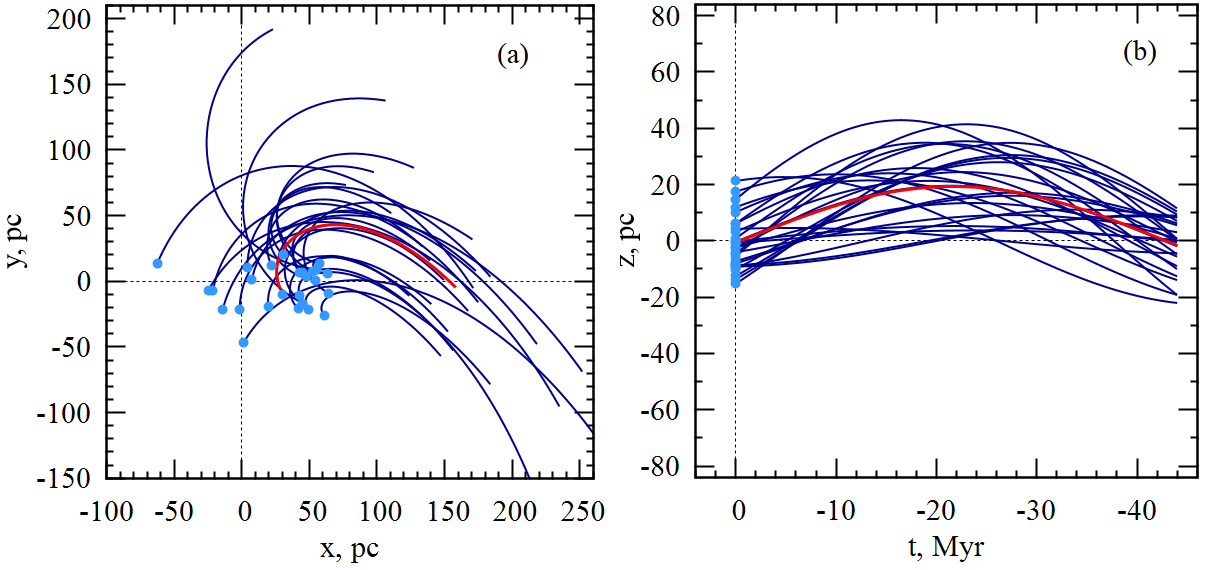}
  \caption{
 (a) The distribution of 27 members of the $\beta$~Pictoris moving group in projection onto the Galactic $xy$ plane and their
traceback trajectories, (b) the vertical distribution of these stars and their traceback trajectories, the trajectories are traced
back on an interval of 45~Myr, the trajectory of the kinematic center of this stellar group is indicated by the red color. We took
all of the data for these stars from the Gaia DR3 catalog.}
 \label{f2-DR3}
\end{center}}
\end{figure}
%%%%%%%%%%%%%%%%%%%%  f2
%%%%%%%%%%%%%%%%%%%%%%%%%%%%%%%%%%%%%%%%%%%%%%%%%%%%%%%%%%%%% F3:
\begin{figure}[t]
{ \begin{center}
   \includegraphics[width=0.9\textwidth]{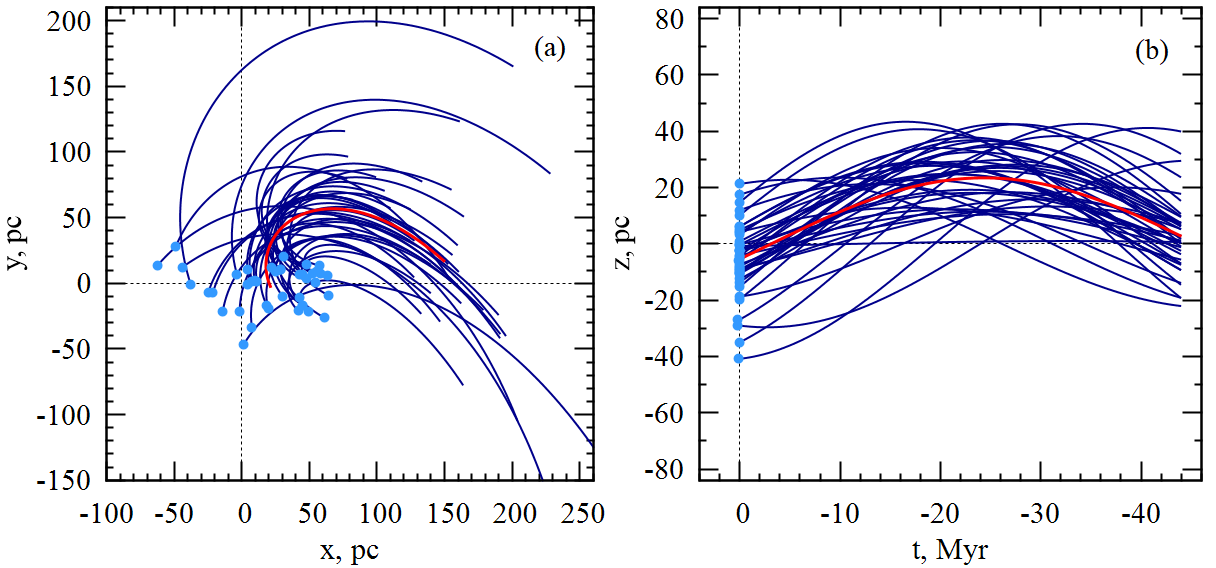}
  \caption{
(a) The distribution of 41 members of the $\beta$~Pictoris moving group in projection onto the Galactic $xy$ plane and their
traceback trajectories, (b) the vertical distribution of these stars and their traceback trajectories, the trajectories are traced
back on an interval of 45~Myr, the trajectory of the kinematic center of this stellar group is indicated by the red color. We took
the line-of-sight velocities of these stars from the SIMBAD electronic database.}
 \label{f3-Simbad}
\end{center}}
\end{figure}
%%%%%%%%%%%%%%%%%%%%  f3
%%%%%%%%%%%%%%%%%%%%%%%%%%%%%%%%%%%%%%%%%%%%%%%%%%%%%%%%%%%%% F4:
\begin{figure}[t]
{ \begin{center}
   \includegraphics[width=0.9\textwidth]{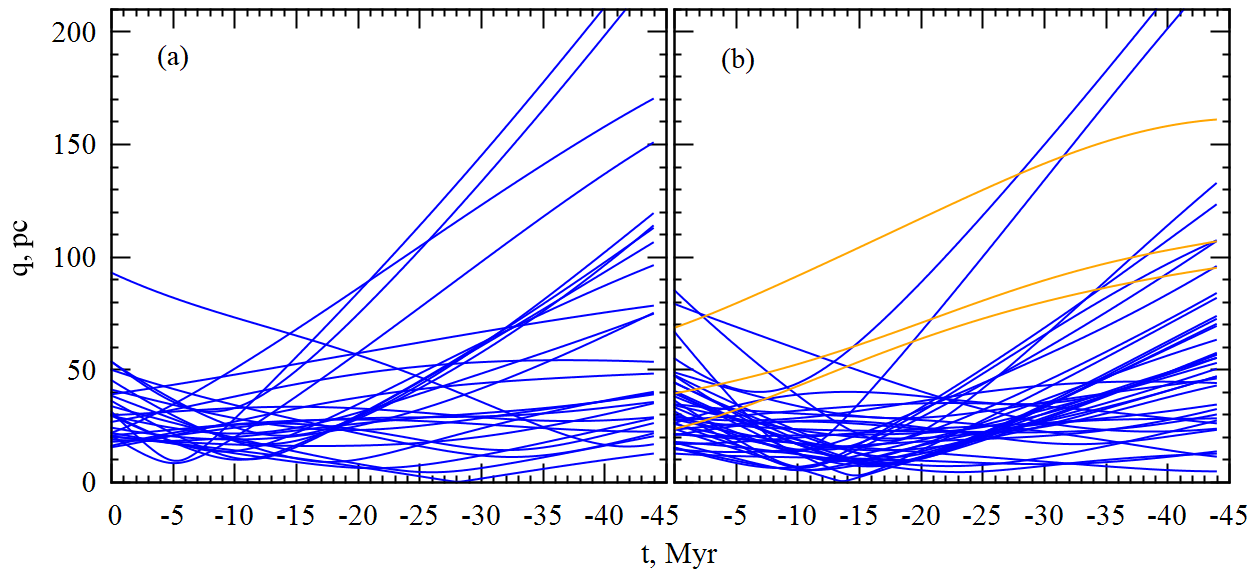}
  \caption{
 Deviations of the stars from the average trajectory (parameter $q$) on the integration interval for 27 stars of the$\beta$~Pictoris
moving group with the line-of-sight velocities from the Gaia DR3 catalog (a) and for 41 stars of this group with the line-ofsight
velocities from the SIMBAD electronic database (b), the yellow color on panel (b) marks the trajectories of the three stars
that were not used in estimating the group age.}
 \label{f4-q}
\end{center}}
\end{figure}
%%%%%%%%%%%%%%%%%%%%  f4
%%%%%%%%%%%%%%%%%%%%%%%%%%%%%%%%%%%%%%%%%%%%%%%%%%%%%%%%%%%%% F5:
\begin{figure}[t]
{ \begin{center}
   \includegraphics[width=0.94\textwidth]{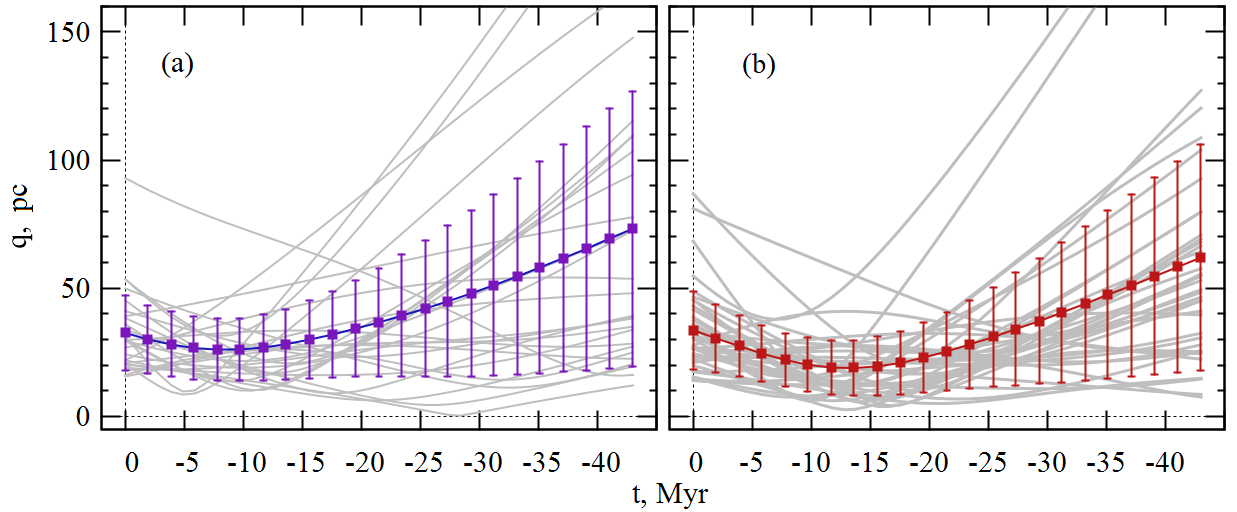}
  \caption{
Deviations of the stars from the average trajectory (parameter $q$) on the integration interval for 27 stars of the $\beta$~Pictoris
moving group with the line-of-sight velocities from the Gaia DR3 catalog (a) and for 38 stars of this group with the line-ofsight
velocities from the SIMBAD electronic database (b), the averaged values with the corresponding dispersions are given.}
 \label{f5-q}
\end{center}}
\end{figure}
%%%%%%%%%%%%%%%%%%%%  f5
%%%%%%%%%%%%%%%%%%%%%%%%%%%%%%%%%%%%%%%%%%%%%%%%%%%%%%%%%%%%% F6:
\begin{figure}[t]
{ \begin{center}
   \includegraphics[width=0.98\textwidth]{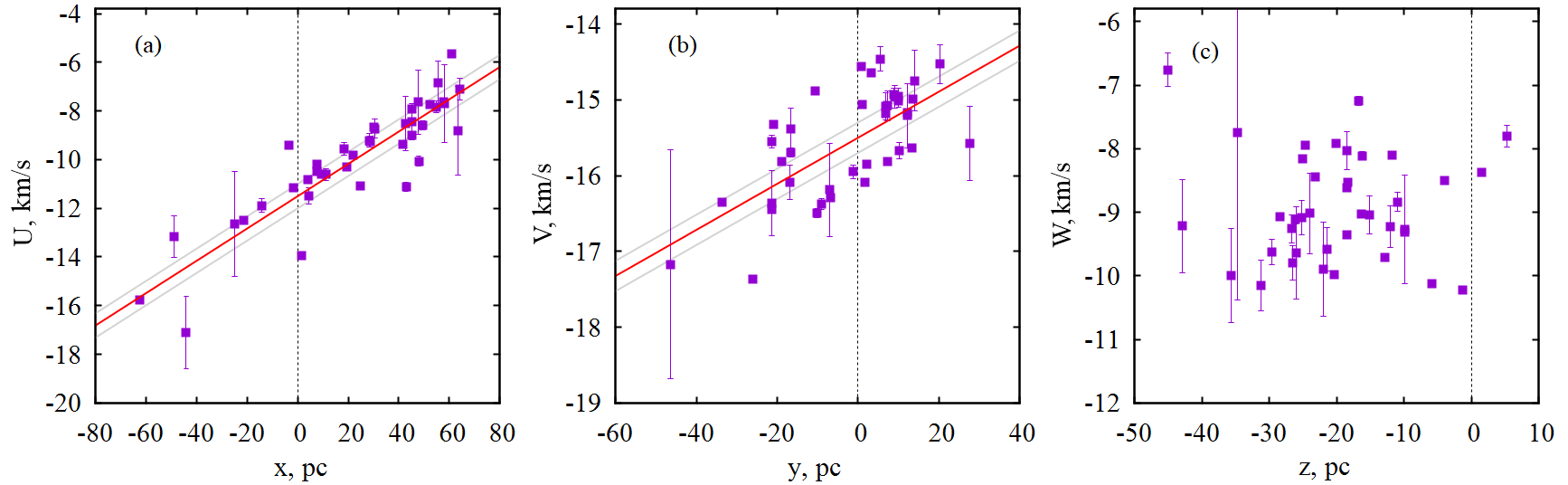}
  \caption{
Velocity $U$ versus coordinate $x$ (a), velocity $V$ versus coordinate $y$ (b), and velocity $W$ versus coordinate $z$ (c) for 38
stars of the $\beta$~Pictoris moving group with the line-of-sight velocities from the SIMBAD electronic database.}
 \label{f6-Keff}
\end{center}}
\end{figure}
%%%%%%%%%%%%%%%%%%%%  f6

\section*{RESULTS}
Figure 1 presents the current positions (cyan circles) and traceback trajectories for 31 stars of the $\beta$~Pictoris moving group. We took the parallaxes, proper motions, and line-of-sight velocities of these stars from the Gaia DR3 catalog. The stellar orbits were integrated into the past on a time interval of 45 Myr. From 27 stars we calculated the average
positions and velocities that we deem the characteristics
of the kinematic center of this stellar group. The four stars whose trajectories are indicated by the orange color in Fig. 1 were discarded due to the large deviations from the average trend. These are the stars with the following Gaia DR3 ID:
 \begin{equation}
 \begin{array}{lll}
 2315849737553379840,\\
 2357025657739386624,\\
 5177677603263978880,\\
 6603693808817829760,
 \label{4-out}
 \end{array}
 \end{equation}
located in the four lower rows of Tables 1 and 2. As can be seen from Table 2, the line-of-sight velocities from the Gaia DR3 catalog differ noticeably from those measured by ground-based methods. Note that these stars are discarded by the 3$\sigma$ criterion when analyzing the dependences of the velocities $U, V,$ and $W$ on the coordinates $x, y,$ and $z,$ as will be discussed below.

Figure 2 presents the current positions and traceback trajectories for 27 stars of the $\beta$~Pictoris moving group together with the trajectory of their kinematic center. Note that the trajectories of the stars were calculated by taking into account the Sun's height above the Galactic plane. Thus, in all our figures the coordinate $z$ reflects the positions of the stars relative to the Galactic plane.

The trajectory of the kinematic center is specified as follows. We calculate the average positions and velocities
of the stellar group ${\overline x}_0,{\overline y}_0,{\overline z}_0$ and ${\overline U}_0,{\overline V}_0,{\overline W}_0$.
Using these values, we construct the trajectory of the kinematic center. Using the coordinate differences (between the star and the kinematic center) $\Delta x,\Delta y,$ and $\Delta z$, we calculate the parameter q for each star at each time of integration in the following form:
 \begin{equation}
 q=\sqrt{\Delta x^2+\Delta y^2+\Delta z^2},
 \label{qq}
 \end{equation}
which characterizes the deviation of the star from the
trajectory of the kinematic center.

Figure 3 presents the current positions and traceback trajectories for 41 stars of the $\beta$~Pictoris moving group. We took the line-of-sight velocities of these stars from the SIMBAD electronic database.

Figure 4 presents the values of the parameter $q$ for both 27 stars of the $\beta$~Pictoris moving group with the line-of-sight velocities from the Gaia DR3 catalog and 41 stars with the line-of-sight velocities from the SIMBAD electronic database. Based on Fig. 4b, we discarded three stars with the following GaiaDR3 ID:
 \begin{equation}
 \begin{array}{lll}
 2357025657739386624,\\
 5177677603263978880,\\
 655168686921108864,
 \label{3-out}
 \end{array}
 \end{equation}
since their trajectories immediately recede greatly from the current position. Thus, not all of the previously discarded four stars from the list (3) are now rejected.

More accurate kinematic data for the three stars from the list (5) may be required. Note that the first two stars in the list (5) have negative values of
the coordinate $z$ with the greatest absolute value (in Fig. 3b their $z\approx-40$~pc). We did not use these three stars to estimate the age of the $\beta$~Pictoris moving group.

Figure 5 actually copies Fig. 4, except that the smoothed averages are given on both panels of the figure. It can be clearly seen that (i) the spatial size of the stellar group 30--40 Myr in the past was much larger than the current one; (ii) there is a minimum of the average line on each panel, although the minimum is deeper in Fig. 5b; and (iii) there is a tendency for the stellar group to expand. Note that in Fig. 5b the stellar orbits were calculated relative to the new trajectory of the kinematic center calculated after the elimination of the outliers. Our analysis of 27 stars with the line-of-sight velocities from the Gaia DR3 catalog (Fig. 5a) gave the following age estimate for the $\beta$~Pictoris moving group:
\begin{equation}
 t=8.5\pm2.2~\hbox {Myr},
 \label{t-DR3}
 \end{equation}
whereas from 38 stars with ground-based line-of-sight velocity determinations (Fig. 5b) we found
 \begin{equation}
 t=13.2\pm1.4~\hbox {Myr}.
 \label{t-Simbad}
 \end{equation}
The errors in the results (6) and (7) were estimated through Monte Carlo simulations. The orbits of the stars were assumed to have been constructed with normally distributed relative errors of 10\%.

Figure 6 presents the following: the dependence of the velocity $U$ on the coordinate $x,$ where the gradient 
$\partial U/\partial x=64.4\pm5.2$~km s$^{-1}$ kpc$^{-1}$ found by the least-squares method based on these data with the boundaries of the confidence intervals is shown; the dependence of the velocity $V$ on $y$ and the gradient $\partial V/\partial y=30.4\pm5.4$~km s$^{-1}$ kpc$^{-1}$; and the dependence of the velocity $W$ on $z.$ In fact, from the data on the three panels presented in the figure we can determine three gradients that in the linear Ogorodnikova--Milne model (Ogorodnikov 1965; Bobylev and Bajkova 2023) are the diagonal elements of the deformation matrix and describe the expansion effects of the stellar system.

To construct Fig. 6, we used 38 stars of the $\beta$~Pictoris moving group with the line-of-sight velocities from the SIMBAD database. There is no dependence of the vertical velocity $W$ on the coordinate $z,$ as can be clearly seen from Fig. 6c. To be more precise, the gradient 
$\partial W/\partial z$ is close to zero, but its error is very large, $\sim$12~km s$^{-1}$ kpc$^{-1}$. Therefore, the volume expansion coefficient of the stellar system ($K_{xyz}=(\partial U/\partial x+\partial V/\partial y+\partial W/\partial z)/3$) cannot be determined reliably.

Based on the gradients  $\partial U/\partial x$ and $\partial V/\partial y$ , we can estimate the plane linear expansion effect of the stellar
system, $K_{xy}=(\partial U/\partial x+\partial V/\partial y)/2$ (the expansion in the $xy$ plane):
\begin{equation}
 K_{xy}=48\pm5~\hbox {km s$^{-1}$ kpc$^{-1}$}
 \label{t-xy}
 \end{equation}
and find the time interval elapsed from the beginning of the expansion of this stellar system to the present day, $t=977.5/K_{xy}$:
\begin{equation}
 t=20\pm2~\hbox {Myr.}
 \label{t-Kxy}
 \end{equation}

\section*{DISCUSSION}
In Lee et al. (2023) the age of the $\beta$~Pictoris moving group was estimated by several methods:
(i) by isochrone fitting and (ii) by lithium-depletion boundary fitting with models that account for the
effect ofmagnetic activity and spots on young, rapidly rotating stars. These authors established that age
estimates are highly model-dependent. Magnetic models with ages of $23\pm8$ and $33^{+9}_{-11}$~Myr provide best fits to the lithium depletion boundary and Gaia $M_G$~ vs. $B_P$--$R_P$ color--magnitude diagram, respectively, whereas a Dartmouth standard model with an age of $11^{+4}_{-3}$~Myr provides a best fit to the 2MASS--Gaia $M_{K_S}$ vs. $B_P$--$R_P$ color--magnitude diagram. Thus, the age estimates for the $\beta$~Pictoris moving group obtained by these authors from the most upto-date data lie in the wide range [11--33]~Myr.

Note the comprehensive summary of age determinations for the $\beta$~Pictoris moving group by various authors presented in Mamajek and Bell (2014). Using isochrone fitting, these authors found $22\pm3$~Myr, while in combination with a lithium abundance analysis they obtained $23\pm3$~Myr. Miret-Roig et al. (2020) also gave a summary of age estimates for the $\beta$~Pictoris moving group covering 20 results obtained from 1999 to 2020. It can be seen from Table 6 of these authors that the estimates lie approximately in the range 10--50 Myr, although they are grouped mainly near 20 Myr.

Our estimate is in good agreement with the kinematic estimates obtained by a number of authors. For example, Crundall et al. (2019) identified ten new probable members and confirmed 48 candidates for members of the $\beta$~Pictoris moving group. Using data
from the Gaia DR2 catalog (Brown et al. 2018) and the line-of-sight velocities from the literature, they estimated the age to be $17.8\pm1.2$~Myr.

Using data from the Gaia DR2 catalog for 81 star, Miret-Roig et al. (2020) obtained a kinematic estimate
of $18.5^{+2.0}_{-2.4}$~Myr. They invoked the line-of-sight velocities of the stars obtained by various authors through ground-based observations.

Couture et al. (2023) obtained a kinematic age estimate of $20.4\pm2.5$~Myr based on data from the Gaia DR3 catalog for 25 members of the
$\beta$~Pictoris moving group.

While analyzing about 40 members of the $\beta$~Pictoris moving group, Torres et al. (2006) found the gradient
$\partial U/\partial x=53$~km s$^{-1}$ kpc$^{-1}$ (without specifying the error), which can be interpreted as a parameter
of the expansion of this cluster along the coordinate $x.$ The value of this parameter found by us is in good
agreement with the estimate by Torres et al. (2006). Our result (8) is more interesting, more physical, and describes the actual distribution of velocity vectors in the $xy$ plane.

On the whole, it can be concluded that there is satisfactory agreement with the kinematic age estimates for the $\beta$~Pictoris moving group obtained in this paper and those of other authors. There is much poorer agrement between the kinematic age
estimates and those obtained by alternative methods in Lee et al. (2024).

\section*{CONCLUSIONS}
Based on data from Lee et al. (2024), we produced two working samples of single stars, probable members
of the $\beta$~Pictoris moving group. The first sample included 31 stars with the trigonometric parallaxes,
proper motions, and line-of-sight velocities from the Gaia DR3 catalog. The second sample included
41 stars with the parallaxes and proper motions from the Gaia DR3 catalog and the line-of-sight velocities
from the SIMBAD electronic database.

To estimate the kinematic age of the $\beta$~Pictoris moving group, we traced back the stellar orbits on an interval of 45 Myr and determined the time when the stellar group had a minimum spatial size. We showed that the kinematic age estimates for the moving group depend strongly on the line-of-sight velocities of the candidate stars used. It can be seen that the line-of-sight velocities of the stars being analyzed derived from ground-based observations (taken from the SIMBAD database) are more reliable than those presented in the Gaia DR3 catalog. They were measured with smaller random errors; their use gives more consistent results between themselves.

As a result, from 38 stars with ground-based lineof-sight velocity determinations we estimated the age by two methods. Both estimates are kinematic. First, our study of the traceback trajectories of the stars gives an estimate of $t = 13.2\pm1.4$~Myr (result (7)). Second, our analysis of the instantaneous stellar velocities suggests an expansion of this stellar system occurring at least in the Galactic $xy$ plane (plane $K$-effect). 
Based on this effect, we found the time interval elapsed from the beginning of the expansion of the $\beta$~Pictoris moving group to the present day, $t = 20 \pm2$~Myr (result (9)).

 \subsubsection*{ACKNOWLEDGMENTS}
We are grateful to the referees for their useful remarks that contributed to an improvement of the paper.

 \subsubsection*{REFERENCES}
 \small

\quad~1. T. A. Agekyan and V. V. Orlov, Sov. Astron. 28, 36 (1984).

2. D. Barrado y Navascu\'es, J. R. Stauffer, I. Song, and J.-P. Caillault, Astrophys. J. 520, L123 (1999).

3. H. Beust, J. Milli, A. Morbidelli, S. Lacour, A.-M. Lagrange, G. Chauvin, M. Bonnefoy, and J. Wang, Astron Astrophys. 683, A89 (2024).

4. V. V. Bobylev and A. T. Bajkova, Astron. Lett. 42, 1 (2016).

5. V. V. Bobylev and A. T. Bajkova, Astron. Lett. 49, 410 (2023).

6. A. G. A. Brown, A. Vallenari, T. Prusti, et al. (Gaia Collab.), Astron. Astrophys. 616, 1 (2018).

7. D. Couture, J. Gagn\'e, and R. Doyon, Astrophys. J. 946, 6 (2023).

8. T. D. Crundall, M. J. Ireland, M. R. Krumholz, Ch. Federrath, M. Zerjal, and J. T. Hansen, Mon.Not. R. Astron. Soc. 489, 3625 (2019).

9. O. Eggen, Mon. Not. R. Astron. Soc. 120, 563 (1960).

10. W. Glieze, Catalog of Nearby Stars (Ver\"off. Astron. Rechen-Inst., Heidelberg, 1969), Vol. 22.

11. J. Holmberg and C. Flinn, Mon. Not. R. Astron. Soc. 352, 440 (2004).

12. L. L. Kiss, A. Mo\'or, T. Szalai, J. Kovacs, D. Bayliss, G. F. Gilmore, O. Bienayme, J. Binney, et al., Mon. Not. R. Astron. Soc. 411, 117 (2011).

13. O. I. Krisanova, V. V. Bobylev, and A. T. Bajkova, Astron. Lett. 46, 370 (2020).

14. R. A. Lee, E. Gaidos, J. van Saders, et al., Mon. Not. R. Astron. Soc. 528, 4760 (2024).

15. B. Lindblad, Ark. Mat., Astron., Fys. A 20 (17) (1927).

16. L. Malo, \'E. Artigau, R. Doyon, D. Lafreniere, L. Albert, and J. Gagne, Astrophys. J. 788, 81 (2014).

17. E. E. Mamajek and C. P. M. Bell, Mon. Not. R. Astron. Soc. 445, 2169 (2014).

18. N. Miret-Roig, P. A. B. Galli,W. Brandner, H. Bouy, D. Barrado, J. Olivares, T. Antoja, M. Romero-Gomez, F. Figueras, and J. Lillo-Box, Astron. Astrophys. 642, A179 (2020).

19. K. F. Ogorodnikov, {\it Dynamics of Stellar Systems} (Fizmatgiz, Moscow, 1965; Oxford, Pergamon, 1965).

20. T. Prusti, J.H. J. de Bruijne, A.G. A. Brown, A. Vallenari, C. Babusiaux, C. A. L. Bailer-Jones, U. Bastian, M. Biermann, et al. (Gaia Collab.), Astron. Astrophys. 595, A1 (2016).

21. A. R. Riedel, C. T. Finch, T. J. Henry, J. P. Subasavage, W.-Ch. Jao, L. Malo, D. R. Rodriguez, R. J. White, et al., Astron. J. 147, 85 (2014).

22. J. E. Schlieder, S. L\'epine, and M. Simon, Astron. J. 140, 119 (2010).

23. J. E. Schlieder, S. L\'epine, and M. Simon, Astron. J. 143, 80 (2012).

24. R. Sch\"onrich, J. Binney, and W. Dehnen, Mon. Not. R. Astron. Soc. 403, 1829 (2010).

25. E. L. Shkolnik, K. N. Allers, A. L. Kraus, M. C. Liu, and L. Flagg, Astron. J. 154, 69 (2017).

26. C. A. O. Torres, G. R. Quast, L. da Silva, R. de La Reza, C. H. F. Melo, and M. Sterzik, Astron. Astrophys. 460, 695 (2006).

27. A. Vallenari, A. G. A. Brown, T. Prusti, et al. (Gaia Collab.), Astron. Astrophys. 674, 1 (2023).

28. B. Zuckerman, I. Song, M. S. Bessell, and R. A.Webb, Astrophys. J. 562, L87 (2001).

29. The HIPPARCOS and Tycho Catalogues, ESA SP--1200 (1997).

\end{document}